\documentclass[journal=acscii, manuscript=article]{achemso}
\SectionNumbersOn

\usepackage{chemformula} 
\usepackage[T1]{fontenc} 

\usepackage{units}
\usepackage{amsmath}
\usepackage{amssymb}
\usepackage{booktabs}
\usepackage{enumerate}
\usepackage{multirow}
\usepackage{tabularx}
\usepackage[symbol]{footmisc}

\usepackage{xr-hyper}
\usepackage[colorlinks=true, allcolors=blue]{hyperref}
\newcommand{\breg}{Bayesian Linear Regression }
\newcommand{\convnet}{Convolutional Neural Network }
\newcommand{\mlpf}{Multi-Layer Perceptron (full) }
\newcommand{\mlpp}{Multi-Layer Perceptron (pca) }

\graphicspath{{./}, {./Graphics/}}

\author{Lo\"ic M. Roch}
\affiliation
{Department of Chemistry and Chemical Biology, Harvard University, Cambridge, MA 02138, USA}

\alsoaffiliation
{Vector Institute for Artificial Intelligence, Toronto, ON M5S 1M1, Canada}

\alsoaffiliation
{Department of Chemistry, University of Toronto, Toronto, ON M5S 3H6, Canada }

\alsoaffiliation
{Department of Computer Science, University of Toronto, Toronto, ON M5S 3H6, Canada }

\author{Semion K. Saikin}
\affiliation
{Department of Chemistry and Chemical Biology, Harvard University, Cambridge, MA 02138, USA}

\alsoaffiliation
{Kebotix, Inc., Cambridge, MA 02139, USA}

\author{Florian H\"ase}
\affiliation
{Department of Chemistry and Chemical Biology, Harvard University, Cambridge, MA 02138, USA}

\alsoaffiliation
{Vector Institute for Artificial Intelligence, Toronto, ON M5S 1M1, Canada}

\alsoaffiliation
{Department of Chemistry, University of Toronto, Toronto, ON M5S 3H6, Canada }

\alsoaffiliation
{Department of Computer Science, University of Toronto, Toronto, ON M5S 3H6, Canada }

\author{Pascal Friederich}
\affiliation
{Department of Chemistry, University of Toronto, Toronto, ON M5S 3H6, Canada }

\alsoaffiliation
{Department of Computer Science, University of Toronto, Toronto, ON M5S 3H6, Canada }

\alsoaffiliation
{Institute of Nanotechnology, Karlsruhe Institute of Technology, 76344 Eggenstein-Leopoldshafen, Germany}

\author{Randall H. Goldsmith}
\affiliation
{Department of Chemistry, University of Wisconsin-Madison, Madison, WI 53706, USA}

\author{Salvador Le\'on}
\affiliation
{Department of Industrial Chemical Engineering and Environment, Universidad Polit\'ecnica de Madrid, 28006 Madrid, Spain}

\author{Al\'an Aspuru-Guzik}
\email{alan@aspuru.com}
\affiliation
{Department of Chemistry, University of Toronto, Toronto, ON M5S 3H6, Canada }

\alsoaffiliation
{Department of Computer Science, University of Toronto, Toronto, ON M5S 3H6, Canada }

\alsoaffiliation
{Vector Institute for Artificial Intelligence, Toronto, ON M5S 1M1, Canada}

\alsoaffiliation
{Lebovic Fellow, Canadian Institute for Advanced Research (CIFAR), Toronto, ON M5G 1M1, Canada \vspace*{0.75cm}}

\title{From absorption spectra to charge transfer in PEDOT nanoaggregates with machine learning \vspace*{0.75cm}}


\begin{document}

\begin{tocentry}

\end{tocentry}

\begin{abstract}
Fast and inexpensive characterization of materials properties is a key element to discover novel functional materials. In this work, we suggest an approach employing three classes of Bayesian machine learning (ML) models to correlate electronic absorption spectra of nanoaggregates with the strength of intermolecular electronic couplings in organic conducting and semiconducting materials. As a specific model system, we consider PEDOT:PSS, a cornerstone material for organic electronic applications, and so analyze the couplings between charged dimers of closely packed PEDOT oligomers that are at the heart of the material's unrivaled conductivity.  We demonstrate that ML algorithms can identify correlations between the coupling strengths and the electronic absorption spectra. We also show that ML models can be trained to be transferable across a broad range of spectral resolutions, and that the electronic couplings can be predicted from the simulated spectra with an 88\% accuracy when ML models are used as classifiers. Although the ML models employed in this study were trained on data generated by a multi-scale computational workflow, they were able to leverage leverage experimental data.
\end{abstract}

\section{Introduction}
Organic-based materials are attractive for optoelectronic device applications, notably due to their low fabrication cost and their relative ease to produce and characterize.\cite{Prineted_book} Not only can the structural properties of these materials be tuned through the functionalization of molecules,\cite{Cheng_ChemRev2009} but they are also composed of elements which are Earth-abundant. In contrast to conventional inorganic electronic materials, organic compounds bring in flexibility,\cite{LOGOTHETIDIS200896} biocompatibility\cite{Rivnay-NatComm:2016} and biodegradability,\cite{Lei5107} as well as self-healing properties.\cite{Oh2016, Ocheje2017} Organic conducting and semiconducting materials hold promises for several application niches, including next-generation wearable and printed photovoltaics,\cite{Facchetti_ChemMat2011,Lipomi_AdvMat2011} fuel cells, \cite{guo2012self,winther2008high} thermoelectrics, \cite{Dubey_JPolSci2011,Wang_AdvEnMat2015,Wei_Materials2015,Bubnova_NatMat2011} and other optoelectronics applications.\cite{sun2015review, Roch2017}\\

One of the fundamental challenges for the design of organic optoelectronics lies in the intrinsic structural disorder of the materials. This disorder occurs on multiple length scales starting from the conformations of single molecules and the nearest-neighbor packing to the formation of multi-molecule domains and nanocrystals. The electronic properties of organic materials are highly sensitive to the packing of composing molecules hence dependent on the processing conditions.\cite{Hinton-JACS:2018} Fast optical probing of local electronic couplings can benefit both applied and fundamental research. On one hand, such a method brings the possibility to combine continuous testing of devices with roll-to-roll device manufacturing technology.\cite{Tabor:2018, MI-report} On the other hand, optical characterization techniques can advance our understanding of charge transport in organic structures. In particular, UV-Vis, XPS, and Raman scattering measurements of thin films of conductive polymers can provide insight on composition and electronic structure, including the nature of charge carriers.\cite{Luo-JMCA:2013, Ouyang-polymer:2004} \\



The microscopic structure of molecule and polymer packing is difficult to measure directly. In contrast, obtaining optical spectra such as IR absorption, Raman scattering, electronic absorption, and fluorescence is more straightforward and requires sufficiently less experimental effort. Therefore, indirect characterization methods play a key role in evaluating the level of disorder. Both optical and electronic transport properties are influenced by the microscopic molecular packing. For example, in the simplest qualitative picture, the close proximity of two molecules yields an overlap of electronic clouds which leads to charge transfer. This proximity also leads to a F{\"o}rster coupling between electronic excitations, which can be observed as changes in the lines in the electronic absorption spectra.\cite{Saikin-Nanophotonics:2013} Moreover, weak charge-transfer excitations are observed if the electronic coupling between molecules becomes sufficient. Because both effects are caused by the molecular interactions, in principle it is possible to find a machine learning model that correlates both of them. The conventional computational approach involves three steps: (i) building physical models that describe both properties of interests; (ii) fitting the parameters of the models to experimental data, e.g., absorption spectra; and (iii) using the fitted models to describe the other property, e.g., conductivity. Such an approach might be tedious, since the interrelations between these properties can be too complex to derive a simple or tractable physical model.\\ 

Herein, we report an alternative approach, where the aforementioned physical model is replaced by machine learning (ML) models. To this end, we design a multi-scale computational workflow where the first three steps -- force-field calculations, molecular dynamic simulations, and quantum-based approaches -- generate the data for the ML algorithms. It is of course possible to employ experimental data in addition if it is available. In this work, and to begin with, we employed ML algorithms to identify correlations between the two properties of interests i.e., strength of intermolecular coupling and electronic absorption spectra. Then, we used the trained ML models as a relative classifier of the coupling strength of a given spectrum with respect to a reference coupling, which is to be defined by a scientist for the application at hand. As a model system demonstrating the reliability of the classifier to identify structures with strong electronic couplings from their electronic absorption spectra, we study pairs of poly(3,4-ethylenedioxythiophene) (PEDOT) oligomers. PEDOT is one of the most technologically-developed conducting polymers. Owing to its high hole conductivity and optical transparency in a doped state\cite{Groenendaal_AdvMat2000} it is has been used as transparent contacts in photovoltaic devices, touch screens, and light-emitting diodes. \cite{Lovenich_PolSci2014} PEDOT-based materials are frequently used as a mixture with polystyrene sulfonate polymers (PEDOT:PSS). In this mixture, PEDOT oligomers transfer the charges while PSS chains play the role of a solid electrolyte. This material becomes conductive at high concentrations of dopant.\cite{PEDOT_book}\\

Although multiple experimental studies have addressed the molecular organization of PEDOT-based materials,\cite{Martin-PR:2010,Groenendaal_AdvMat2000, Takano_MacroM2012, Nardes-AM:2007, Crispin-JPS:2003, Kemerink-JPCB:2004, VanDeRuit-AFM:2013, Crispin-ChemMater:2006, Luo-JMCA:2013, Ouyang-polymer:2004} the microscopic electronic states that lead to high conductance and the interplay between these states, optical properties, and the material structure have yet to be determined. As a matter of fact, the key factor for practical applications of PEDOT-based materials lies in understanding the relations between their solid-state packing and their unique electronic properties. Consequently, the main obstacle to elucidating this relationship is the strong structural disorder that appears on multiple length scales and is highly sensitive to the thin film preparation procedure.\cite{Groenendaal_AdvMat2000}\\

Hereafter, we demonstrate that our ML models 
confirm the existence of correlation between the coupling strengths and the electronic absorption spectra. We also show the robustness of our ML models with respect to potential spurious statistical correlations to capture the relevant physical correlations. Finally, we use the ML models as classifiers to determine whether a given electronic absorption spectrum of interest relates to a coupling strength above or below an \textit{a priori} selected reference coupling strength. Such an approach has proven to be reliable and robust, reaching an average error rate of only 12\% when employing a Bayesian convolutional neural network. The importance of such a classifier becomes apparent in the context of the self-driving laboratories,\cite{SciRobot18:Roch,Tabor:2018,Hase-2019,MI-report,Dimitrov-ACS:2019} where the goal of the experimentation process to identify fabrication procedures for aggregates yielding high conductivities is embodied as an optimization procedure. \\

\section{Methods}
\label{sec:methods}
\begin{figure}[htb!]
\centering
\includegraphics[width=0.9\textwidth]{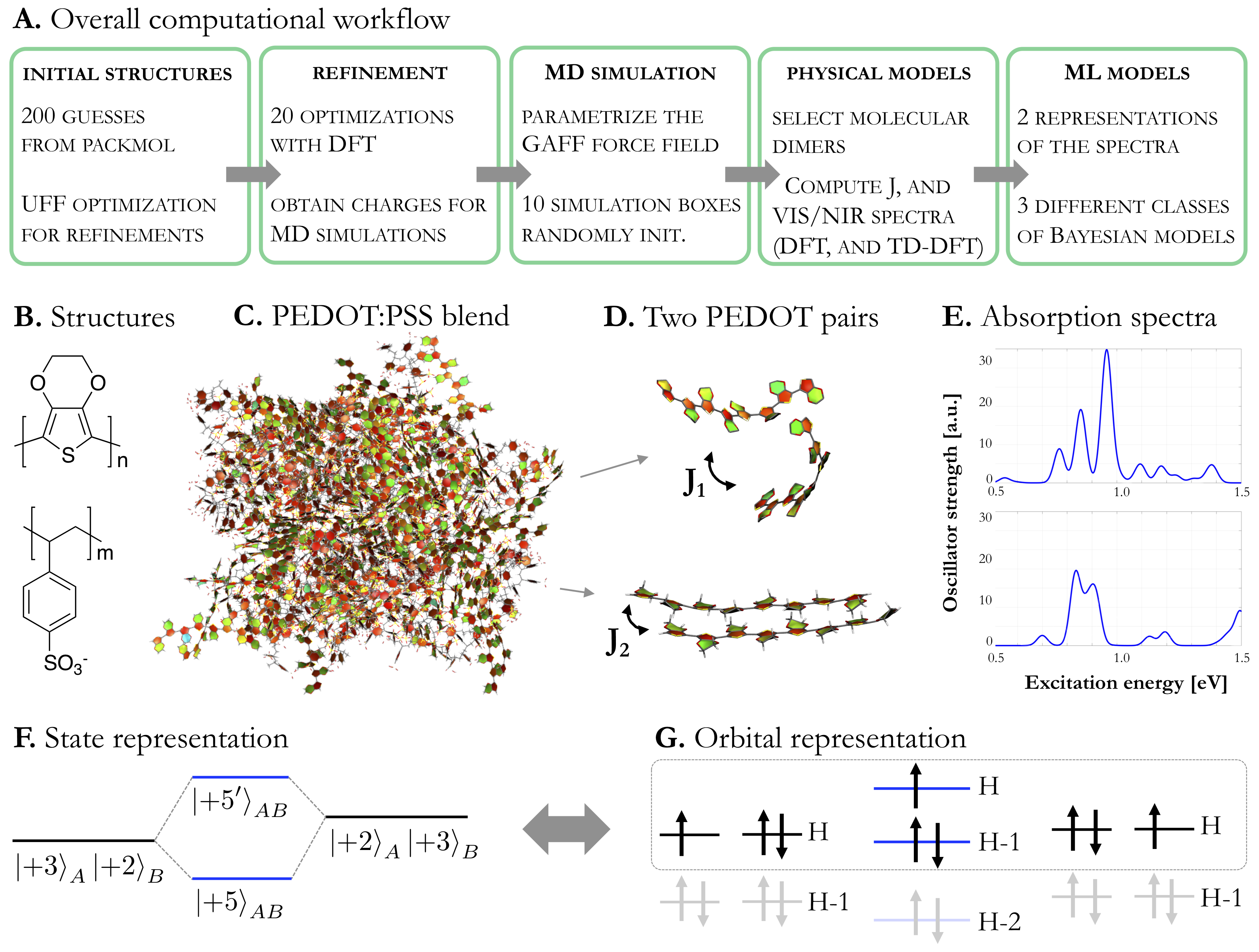}
\caption{The computational pipeline used, from structure generation to correlating $J$ and electronic absorption spectra. (A) General workflow highlighting the steps and summarizing the methods involved. (B) Structures of PEDOT and PSS, represented in the top and bottom panels, respectively. (C) One of the ten supercells of PEDOT:PSS blends. (D) Two distinct example pairs of PEDOT oligomers extracted from the PEDOT:PSS bulk. $J_1$ and $J_2$ are the coupling strengths for each of the pairs. (E) Associated simulated electronic absorption spectra. (F) State, and (G) orbital representation of the monomers and dimers involved in the calculation of the coupling strength, $J$.}
\label{fig:concept}
\end{figure}

This section details the computational workflow designed to generate the data and to correlate electrical and optical properties using ML algorithms. The workflow is depicted in Fig.~\ref{fig:concept}. Each of the five composing steps (i.e., initial structures, refinement, MD simulation, physical models, ML models in Fig.~\ref{fig:concept}A) are described in their corresponding subsections.\\

Hereafter, we assume that the packing of conjugated oligomers in the solid PEDOT:PSS mixture depends on the initial preparation procedure and post-processing steps. For PEDOT:PSS, such post-processing steps, typically evaluated via trial and error, are critical to achieve peak performance. Furthermore, it is hypothesized that PEDOT:PSS solid films consist of grains with a hydrophobic and highly conductive PEDOT-rich core and a hydrophilic insulating PSS-rich shell.\cite{David-2016,Takano_MacroM2012} This phase segregation of PEDOT and PSS occurs on a scale beyond current computational capabilities and, thus, is not captured by our model. Nonethless, our computational workflow allows to study the disorder within each of these grains.\\

\paragraph*{Initial structures.} To screen the orientation stability of the PEDOT:PSS complex within the grains, 100 starting structures were generated using the Packmol software package.\cite{Packmol2009} Each of the complexes consisted of one PEDOT chain with eight 3,4-ethylenedioxythiophene units ($n=8$, Fig.~\ref{fig:concept}B) carrying two positive charges, and two PSS chains consisting of three poylstyrene sulfonate units ($m=3$, Fig.~\ref{fig:concept}B) with one negative SO$_3^-$ and two SO$_3$H groups per chain. In the generation of the initial structures, we constrained the SO$_3^-$ group of the PSS chain to point towards the positive PEDOT chain. These initial structures were then optimized using a classical force field (FF) approach.

\paragraph*{Refinement.} The 20 energetically most stable PEDOT:PSS complexes obtained from the initial structure search were relaxed at the B97-D/6-31G(d,p) level of theory in the gas phase, using the Gaussian software package.\cite{g09} Note that the influence of the solvent was found to be negligible, and that the performance of the B97-D functional on geometries has already been assessed in previous work.\cite{Roch:jctc-2017} Single point energy calculations at the HF/6-31G(d,p)//B97-D/6-31G(d) level were performed on the 20 relaxed complexes to parametrize the charges of the molecular dynamics simulation, as customary with GAFF.\cite{GAFF2004}

\paragraph*{Molecular dynamics simulation.} Ten molecular dynamics simulations were carried out on PEDOT:PSS model systems in periodic cubic boxes of size\footnote{Note: average size of the different boxes for the duration of the NPT simulations} \unit[61.32]{\AA} $\times$ \unit[61.32]{\AA} $\times$ \unit[61.32]{\AA} for a density of \textit{ca.} \unit[1.4]{g/cm$^3$}, with the LAMMPS software package.\cite{PLIMPTON19951} The GAFF force-field was chosen to describe the systems. PEDOT oligomers with eight repeat units ($n=8$, Fig.~\ref{fig:concept}B) and a +2 charge were considered, while the PSS atactic chains consisted of 20 repeat units ($m=20$, Fig.~\ref{fig:concept}B), with four deprotonated units randomly distributed in the sequence of each chain. Note that the PSS chain length was increased from $m=3$ to $m=20$ to better represent experimental blends. Additional details can be found in the Supporting Information (see section \ref{sec:si_md}).

\paragraph*{Physical models to compute $J$, and simulate the electronic absorption spectra.}
For two interacting PEDOT monomers, denoted $A$ and $B$ from hereon, the strength of the charge transfer integral $J$ can be determined in several ways.\cite{Bredas_JACS1983, Bredas_CPL2002, Bredas_JACS2008, martin2000monodisperse, bredas:jcp-2007, Apperloo_ChemEurJ2002, Beljonne_AdvFM2001, Li-JCP:2007, deibel:prb-2011, Kubas-JCP:2014, Cave-JCP:97} Bi-polaron charge transport models have previously been discussed in the case of PEDOT systems.\cite{Bredas_AChR1985} For the sake of simplicity, we assume a single-polaron transport model. Nonetheless, the ML models used for electronic coupling prediction are agnostic to the type of transport and will learn correlations between $J$ and electronic absorption spectra independently of the type of the charge transport model. To calculate the coupling $J$, we used the framework of a tight-binding formalism\cite{MolCrys-Book,Bassani-Book} as well as a Kohn-Sham orbital based method.\cite{bredas:jcp-2007,deibel:prb-2011} In both cases, orbital energies were obtained at the B3LYP/def2-SV(P) level of theory. The choice of basis set and functional balances computational cost and accuracy. Detailed results on the performance of the def2-SV(P) results can be found in the Supporting Information (see section \ref{sec:basis_set}). \\

The nearest-neighbor PEDOT dimers were extracted across the ten simulation boxes. These pairs were selected according to a distance criterion; any pair of PEDOT molecules having at least two heavy (i.e., non hydrogen) atoms at a distance closer than \unit[4]{\AA} is selected. The cutoff distance was taken to be comparable to the sum of the van der Waals radii of these heavy atoms. This procedure lead to the selection of 1,420 PEDOT pairs, ulteriorly used to model the strength of the charge transfer, and to simulate the electronic absorption spectra.\\

Tight-binding formalism is based on the change of orbital energies when going from isolated monomers to dimer systems \cite{Bredas-ChemRev:2004}

\begin{equation}
\centering
J=\sqrt[]{\left(\Delta\epsilon_{\left|AB\right>,+5}^{H,L}\right)^2 - \frac{1}{4}\left[ \left(\epsilon_{\left|A\right>,+3}^{H,H-1} + \epsilon_{\left|B\right>,+2}^{H,H-1}\right) - \left(\epsilon_{\left|A\right>,+2}^{H,H-1} + \epsilon_{\left|B\right>,+3}^{H,H-1}\right) \right]^2},
\label{eq:j}
\end{equation}

\noindent where $\Delta\epsilon_{\left|AB\right>,+5}^{H,L}$ is the splitting between the HOMO and the HOMO-1 level of the dimer $AB$ with the charge $q=+5$, and $\epsilon_{\left|i\right>,+q}^H$ is the HOMO energy of monomer $i$, carrying charge $+q$. Note that the correction due to the offset between the HOMOs of the monomers is negligible; hence, $J$ is mostly governed by the splitting $\Delta\epsilon_{\left|AB\right>,+5}^{H,H-1}$. This formalism considers frontier orbitals assuming that only highest occupied orbitals are hybridized due to the electronic coupling between the oligomers (see Fig.~\ref{fig:concept}F-G). Both the presence of non-equilibrated charges and non-zero spin makes the model rather complicated. Nonetheless, our interpretation of  eq.~\ref{eq:j} is a lower estimate for the electronic coupling between the oligomers. Note that this model can also be used to describe bi-polaron transport.\\

The second approach to calculate charge transfer integrals uses the Kohn-Sham orbitals of isolated monomers as well as the Fock matrix and the overlap matrix of the dimer systems:\cite{bredas:jcp-2007,deibel:prb-2011}

\begin{equation}
J_{\text{AB}}=\frac{F_{\text{AB}}-\frac{1}{2}(F_{\text{AA}}+F_{\text{BB}})S_{\text{AB}}}{1-S_{\text{AB}}^2}
\label{eq:j2}
\end{equation}

\noindent The matrix elements $F_{\text{AB}}=\left<\text{A}|F_{\text{dimer}}|\text{B}\right>$ and $S_{\text{AB}}=\left<\text{A}|S_{\text{dimer}}|\text{B}\right>$ are calculated using the Fock and overlap matrices of a molecular dimer system with a charge of $+4$. The states $\left.|\text{A}\right>$ and $\left.|\text{B}\right>$ are the highest occupied molecular orbitals of the doubly positively charged monomers. The distributions of electronic couplings obtained with eq. \ref{eq:j}, and with eq. \ref{eq:j2} are depicted in Fig.~\ref{sec:si_coupling_strength_distributions}.\\

The electronic absorption spectra of the 1,420 PEDOT pairs were simulated using the computed TD-CAM-B3LYP/def2-SV(P) transitions, in the gas phase. Note that for the simulation of the spectra, the influence of solvation was found to be negligible. We employed a Lorentzian distribution to broaden the TD-DFT transitions. This translates the discrete oscillator strength, $f$, and transition energies, $\omega$, to a continuous spectra to resemble experimental outcomes. The broadening was chosen to be \unit[50]{meV}.\cite{Randall:NL-2018}\\

Some insights about correlations between electronic absorption spectra and intermolecular charge transfer can be obtained only for the case when the coupling is weak. A completely relaxed doubly charged oligomer composed of eight to ten units would have a strong electronic transition at about 0.9 -- \unit[1.0]{eV}.\cite{David-2016} This transition is predominately composed of HOMO and LUMO orbitals. For an oligomer with an odd number of charges, additional HOMO-1 $\rightarrow$ HOMO transition appears at lower frequencies, at \textit{ca.} \unit[0.5]{eV}.  Therefore, the low frequency part of the spectra of dimers without the interaction should be composed of three lines -- a low-frequency, weak transition and a strong doublet. Electronic or excitonic interaction between the molecules modulate the spectra. Specifically, weak charge transfer transitions should appear at the low frequency tail of the spectra. The intensities of these transitions should be sensitive to the coupling strength, while their frequencies should be more stable as determined by the alignment of the molecular energy levels. However, this intuitive picture fails for intermediate and strong intermolecular couplings. In the latter case, the intramolecular states hybridize, which in turn leads to a realignment of the energy levels and a redistribution of the oscillator strengths among multiple transitions. Yet another advantage of our ML approach is that it allows to capture correlations between the electronic interaction and optical spectra independently of the coupling regime used as our ML independent variable.

\paragraph*{Machine learning models to identify correlations between electronic absorption spectra and coupling strengths.} Correlations between electronic absorption spectra and coupling strengths were identified with ML models at different levels of complexity. To mimic experimental conditions, we encoded the Lorentzian broadened electronic absorption spectra based on their intensities at specific frequencies, using a \unit[1]{meV} binning on the considered frequency domain (\unit[0.5]{eV} to \unit[1.5]{eV}). A total of 170 (\unit[11.97]{\%}) of the 1,420 spectra-coupling pairs were randomly selected to construct a test set. The remaining 1,250 (\unit[88.03]{\%}) of the dataset were used for ten-fold cross-validation. The size of the dataset motivates the use of Bayesian models for a robust and transferable identification of relevant physical correlations. \\
    
\begin{figure}
        \centering
        \includegraphics[width=0.75\textwidth]{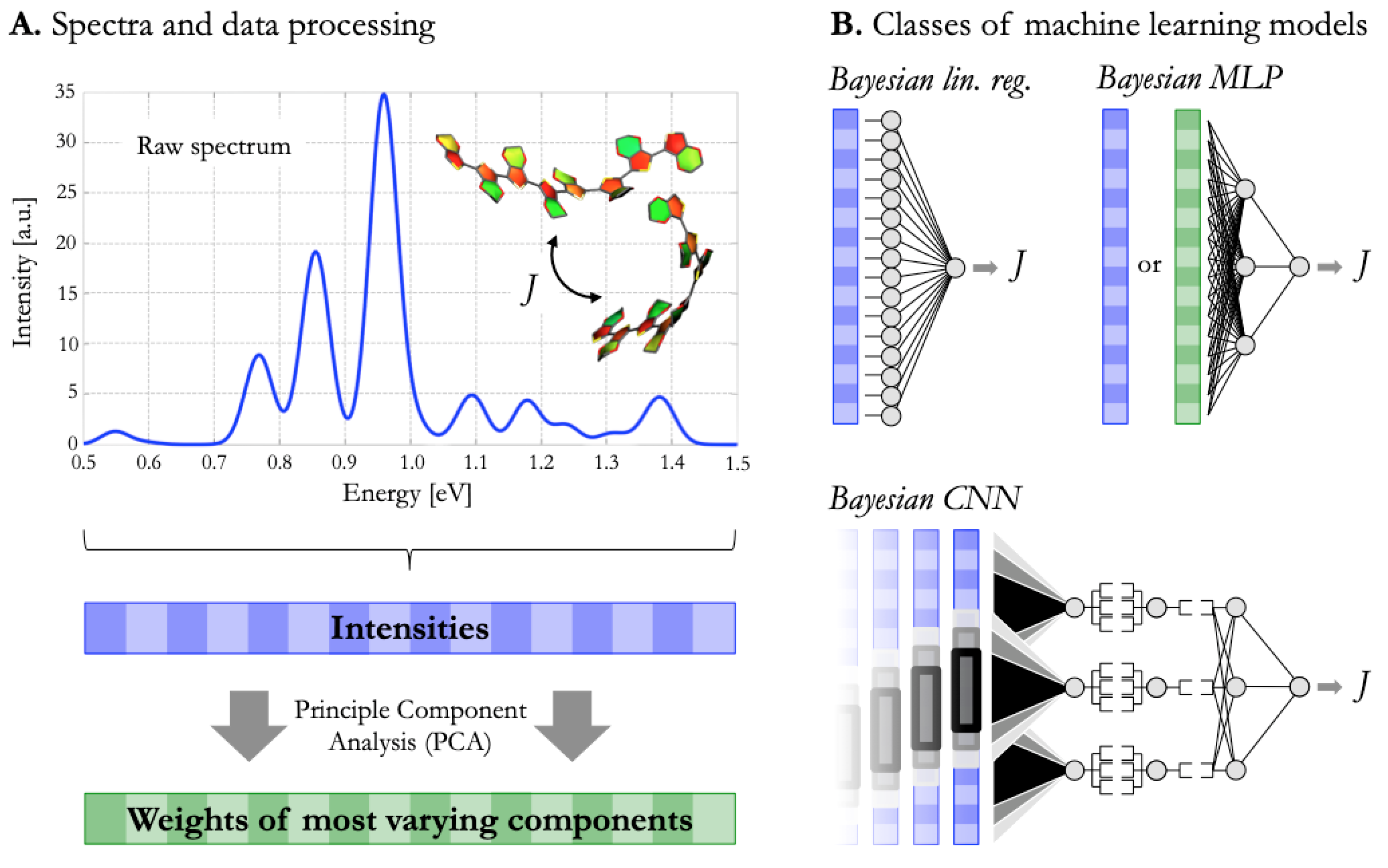}
        \caption{Schematic representation of (A) the data processing and (B) the three classes of ML models, depicting Bayesian linear regression models, Bayesian multi-layer perceptrons (MLPs), and Bayesian one-dimensional convolutional neural networks (CNN). Note that the MLPs are trained either on the raw intensities (blue array) or on the compressed representation after a principle component analysis (green array).}
        \label{fig:ml_models}
    \end{figure}

    Specifically, we employed three different classes of Bayesian models (see Fig.~\ref{fig:ml_models}): (i) Bayesian linear regression models assume a linear dependence of coupling strengths on electronic spectra and thus present the simplest approximation; (ii) Bayesian multi-layer perceptrons (MLPs) are Bayesian generalizations of conventional deterministic MLPs with similar flexibility to model non-linear relations while retaining the robustness to overfitting of Bayesian methods; and (iii) Bayesian one-dimensional convolutional neural networks (CNNs) are special cases of MLPs which have the potential to efficiently exploit spatial correlations in the presented features due to their sparse nature. All models were set up to predict coupling strengths directly from the intensities of the associated electronic absorption spectra at different frequencies. Additionally, we constructed Bayesian MLPs which are trained on a compressed representation of the electronic absorption spectra obtained from principle component analysis (PCA). All models were trained based on an early-stopping criterion. Hyperparameters for all four models are optimized in a random grid search, and the details are provided in the Supporting Information (see section \ref{sec:si_hyperparameter_optimization}).
    
\paragraph*{Machine learning models as relative classifiers.} The aforementioned ML models trained for predicting absolute values of coupling strengths from electronic absorption spectra can be used to classify the conductivity performance associated with the electronic absorption spectra of the materials. Instead of asking for the absolute value of the coupling strength, the model provides an estimate for whether the considered coupling is above or below a reference. This reference is a hyperparameter defined by a scientist as a threshold for high and low values of $J$.\\

\section{Results and discussion}
\begin{figure}[!ht]
        \centering
        \includegraphics[width = 0.75\textwidth]{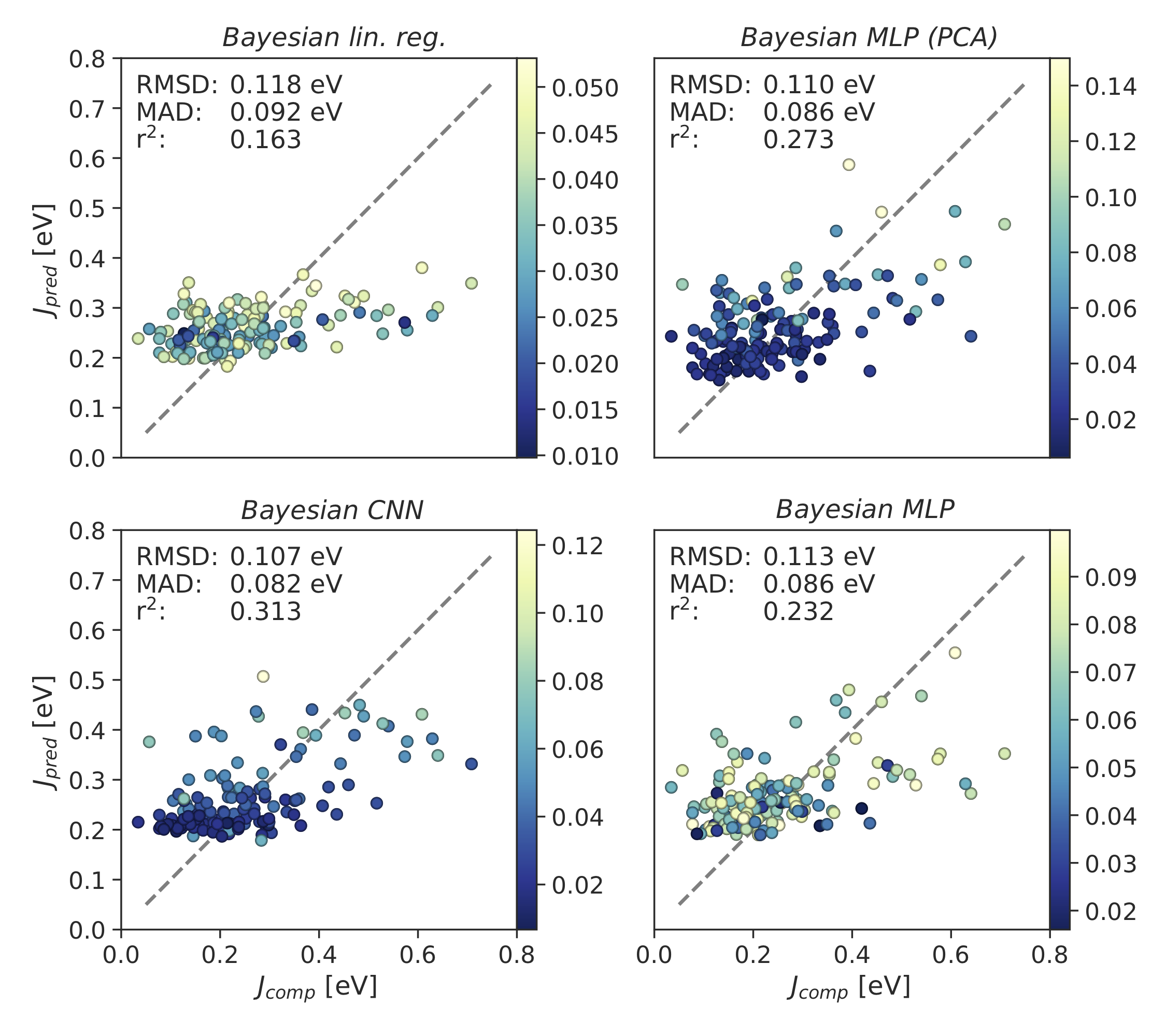}
        \caption{Coupling strengths predicted by the employed machine learning models in comparison to coupling strengths obtained with the tight-binding formalism. All reported predictions are shown for the test set, and were obtained from averaging $200$ prediction samples from each of the models. Prediction uncertainties are color-coded. We report three comparative metrics to assess the prediction accuracies of each model: root-mean-square deviation (RMSD), mean-absolute deviation (MAD), and coefficients of determination ($r^2$). Based on all three metrics, Bayesian CNNs provide the most accurate predictions.}
        \label{fig:test_set_predictions}
    \end{figure}

We begin by discussing our results with the performance of the ML models to identify correlations and to predict absolute values of coupling strengths from electronic absorption spectra. Fig.~\ref{fig:test_set_predictions} illustrates the accuracies of all four models to predict coupling strengths computed from the tight-binding formalism with all four models after a full hyperparameter optimization. Then, we detail the test designed to assess the performance of the ML models to capture relevant physical correlation. We continue our discussion with the results obtained when the ML models are used as relative classifier, where associated error rates are reported in Fig.~\ref{fig:contingency_plot}. We also highlight the practicality of such an approach in discovery applications with the self-driving laboratories. Finally, we discuss the robustness of our ML models upon variations of the peak broadening, which experimentally translates to noise. Fig.~\ref{fig:predictions_for_broadenings} illustrates the performance of our ML models at different broadening.
    
    \subsection{Correlation between electronic spectra and coupling strengths}
    All constructed ML models are able to predict coupling strengths at positive coefficients of determination ($r^2$), which indicate that the coupling strengths correlate with the electronic absorption spectra of the PEDOT dimers and that the models are capable of identifying this correlation. Bayesian CNNs provide the most accurate predictions based on all computed comparative metrics ($r^2 = 0.313$, RMSD $= 0.107$ eV, MAD $= 0.082$ eV)\footnote{RMSD: root-mean-square deviation, MAD: mean-absolute deviation, and $r^2$: coefficients of determination} and Bayesian linear regression yields the least accurate predictions ($r^2 = 0.163$, RMSD $= 0.118$ eV, MAD $= 0.092$ eV). We further observe an improved prediction accuracy when compressing the electronic absorption spectra via PCA for the Bayesian MLP models. \\
    
    Despite the relatively small size of the dataset, we found that the studied ML models, most notably the Bayesian CNN, present an efficient approach to identify the relevant correlations. Estimations of the sampling efficiency of the Bayesian CNN model suggest that it can be trained to reach similar prediction accuracies with only 850 instead of 1,250 training points. No significant improvement in the prediction accuracy is observed when increasing the size of the training set from beyond 850 to 1,250 examples (see Supporting Information, section \ref{sec:si_sampling_efficiency}). This observation, in conjunction with the generalization of the models observed for the test set predictions, indicates that the Bayesian CNN exploits all identifiable correlations to their full extent. \\
    
    
	To ensure that our ML models did not capture spurious correlations that could arise from the methods and formalisms employed to compute the electronic absorption spectra and model the coupling strengths, we tested for the nature of the identified correlations by training the Bayesian CNNs to predict couplings strengths obtained with the Kohn-Sham orbital formalism from the same electronic absorption spectra. Details are provided in the Supporting Information (see section \ref{sec:si_cross_predictions}). The trained Bayesian CNNs achieve prediction accuracies of $r^2 = 0.264$ on the same test set. In addition, we constructed a hybrid dataset, where half of the couplings are randomly chosen from the tight-binding formalism and the other half from the Kohn-Sham orbital formalism. Again, the trained Bayesian CNNs achieve prediction accuracies of $r^2 = 0.280$ indicating that the presented ML models do not capture potential spurious statistical correlations but extract the relevant physical correlations. \\ 
    
    \subsection{Machine learning models as relative classifiers}
    The practicality of the presented trained ML models becomes more apparent when weakening the requirement of accurate absolute predictions of the coupling strengths to accurate relative predictions, which are of interest in discovery applications. Rather than requesting an estimate for the actual numeric value of the coupling strength, the trained model is used to determine if a given electronic absorption spectrum of interest relates to a coupling strength above or below an \textit{a priori} selected reference coupling strength. For such scenarios, the prediction accuracy of the model can be assessed by treating it as a binary classifier to determine true positive and true negative rates for different reference coupling strengths.\\
    
\begin{figure}[!ht]
            \centering
            \includegraphics[width = 0.75\textwidth]{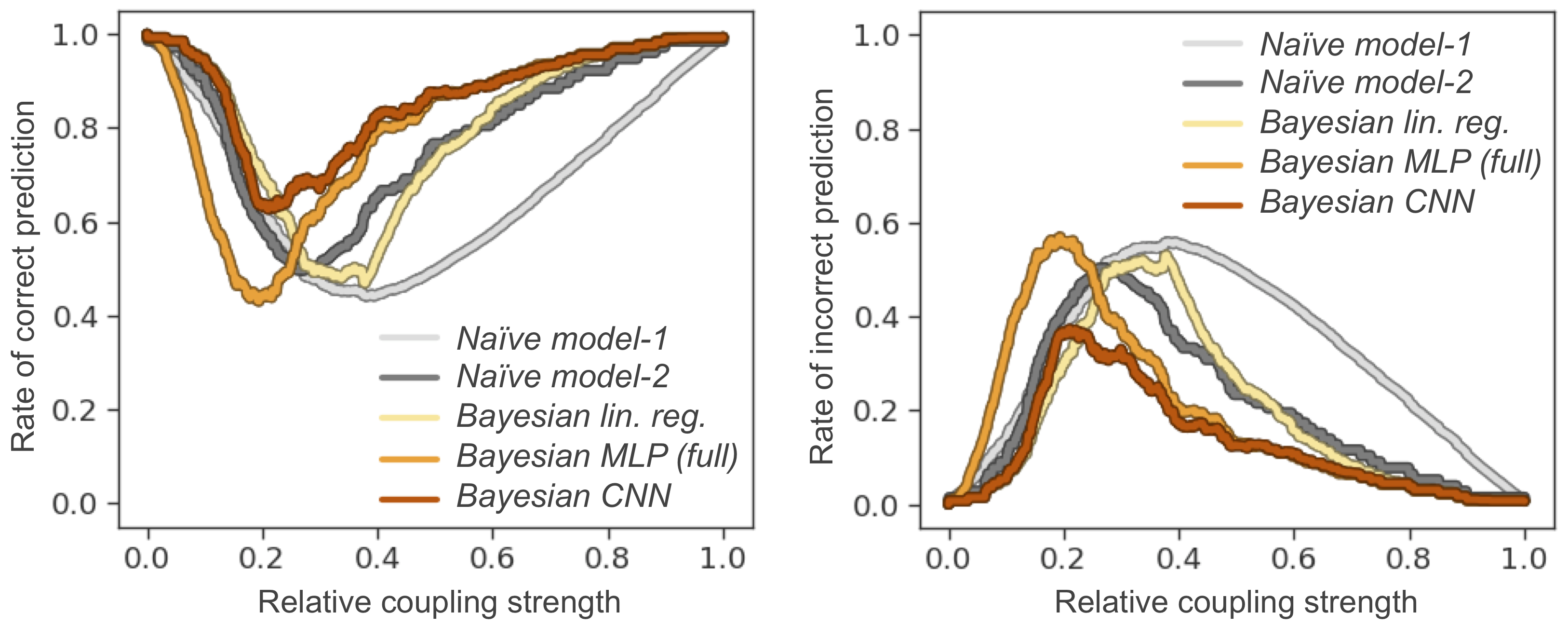}
            \caption{Probabilities to make correct predictions for $J$ regarding the order with respect to another reference $J$ (left panel) vs. making incorrect predictions (right panel).}
            \label{fig:contingency_plot}
        \end{figure}

We assess the prediction accuracies of such relative classifiers by estimating the probability of the model to make a correct prediction, i.e., predicting the coupling to be above the reference when it is above or predicting the coupling to be below the reference when it is below, versus the probability of the model to make an incorrect prediction, i.e., predicting the coupling to be above the reference when it is below and vice versa (see Fig.~\ref{fig:contingency_plot}). These probabilities are estimated for different reference coupling strengths, spanning the entire range of coupling strengths computed with the tight-binding formalism. \\
    
        Fig.~\ref{fig:contingency_plot} illustrates the error rates, i.e., wrongly predicting a coupling strength to be above or below the considered reference coupling strength, for the trained ML models along with two na\"ive models for comparison. The most na\"ive model draws random samples from a uniform distribution to predict the coupling strength for a given electronic absorption spectrum (model-1, depicted in light grey in Fig.~\ref{fig:contingency_plot}). A slightly more sophisticated na\"ive model predicts by drawing random samples from the distribution of coupling strengths (model-2, depicted in dark grey in Fig.~\ref{fig:contingency_plot}).\\
    
	 We find that the error rates for the two na\"ive models yield the largest error rates: \unit[31.9]{\%} and \unit[20.2]{\%}. Bayesian linear regression scored an average error rate of \unit[19.0]{\%}, which, despite its simplicity, already provides an advantage over simple models and captures some of the relevant correlation in the dataset. The lowest error rate is observed for the Bayesian CNN with a \unit[12.6]{\%} average error for coupling strengths chosen within the range of smallest and largest computed coupling strengths. Additionally, it is noted that Bayesian CNN never exceeds an error of \unit[38]{\%} for any chosen reference coupling. In fact, if the focus of discovery process is to identify fabrication procedures and post-processing steps leading to large coupling strengths (above \unit[0.6]{eV}), the Bayesian CNN yields error rates of less than \unit[10]{\%}. We suggest that the trained ML models can be applied to classify the coupling strengths with respect to a reference coupling strength with reasonable confidence.\\

    \subsection{Robustness of the machine learning models}
\begin{figure}[!ht]
        \centering
        \includegraphics[width = 0.75\textwidth]{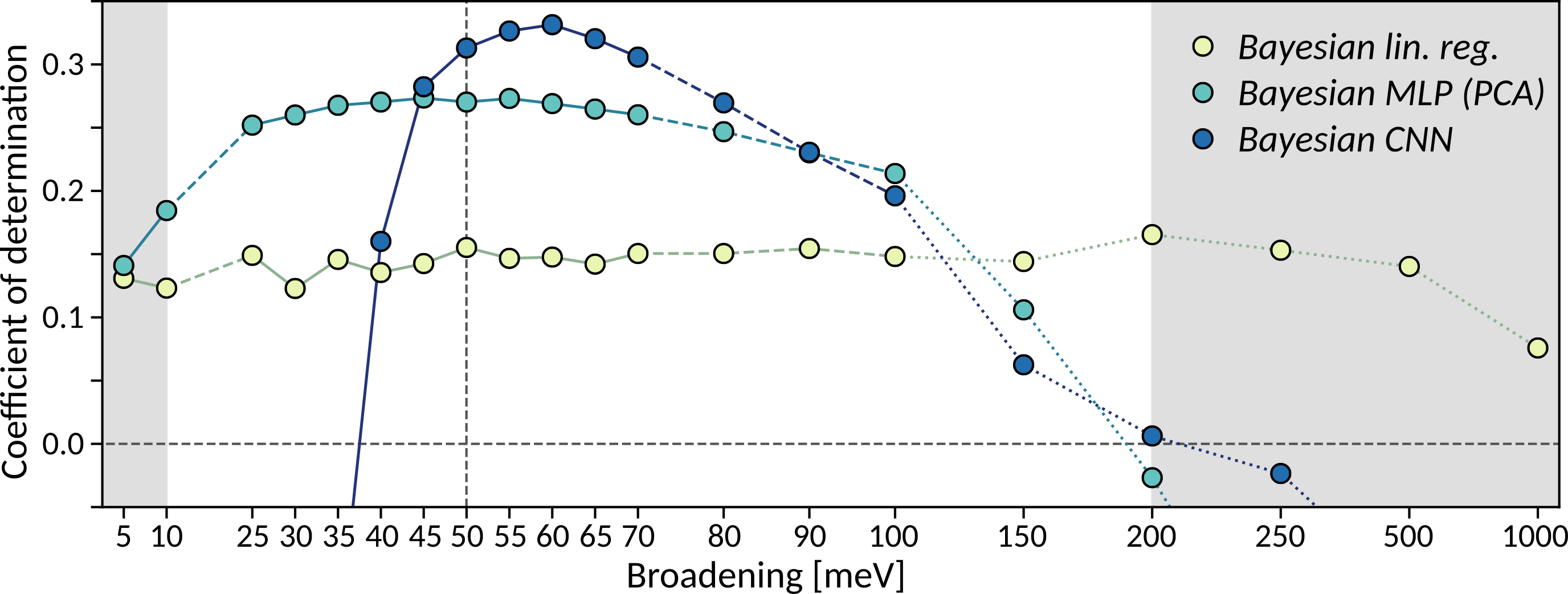}
        \caption{Prediction accuracies measured with $r^2$ coefficients for all models when predicting couplings from spectra at different broadenings. All three models depicted have been trained on a spectra at a \unit[50]{meV} broadening (dashed vertical line). The regime ranging from \unit[10]{meV} to \unit[200]{meV} (highlighted in white) corresponds to the expected experimental resolution: while a \unit[10]{meV} broadening is too small for a room temperature measurement, a \unit[200]{meV} broadening corresponds to an unphysically fast dephasing rate. Note that the horizontal axis, \textit{Broadening [meV]}, is not linear.}
        \label{fig:predictions_for_broadenings}
    \end{figure}

    Finally, we estimate the dependence of the model performances on the particular choice of the peak broadening. While we demonstrated that for this particular choice, the trained ML models are indeed capable of identifying the relevant correlations between the electronic absorption spectra and the coupling strengths, experimentally obtained electronic absorption spectra might be noisy and feature peaks at slightly varying broadenings. The robustness of the model predictions for different broadenings is tested by predicting coupling strengths from electronic absorption spectra at different broadenings, ranging from \unit[5]{meV} to \unit[1,000]{meV}. Note that while a \unit[10]{meV} broadening is too small for a room temperature measurement, a \unit[200]{meV} broadening would correspond to an unphysically fast dephasing rate. Fig.~\ref{fig:predictions_for_broadenings} summarizes the coefficients of determination for predictions of coupling strengths from electronic absorption spectra generated at different broadenings. Note that all coupling strengths are predicted by models, which were trained on electronic absorption spectra at a \unit[50]{meV} broadening. 
    
    We find that prediction accuracies of the Bayesian linear regression model are mostly insensitive to the particular broadening value (yellow trace). Only for very small broadenings below \unit[10]{meV} and very large broadenings above \unit[200]{meV} degradations in the predictive power can be observed. Bayesian MLPs (turquoise trace) show faster degradations in their prediction accuracy for small and large broadenings, but maintain comparative predictive powers across broadenings of \unit[25]{meV} to \unit[100]{meV}. Bayesian CNNs (blue trace) are the least robust with respect to changes in the broadening, with accurate predictions only within the \unit[45]{meV} to \unit[80]{meV} interval. Nevertheless, they demonstrate their predictive power for varying broadenings despite having been trained on peaks of one particular broadening, indicating that ML models can indeed be trained to identify relevant correlations without being overly sensitive to the broadening of the peaks.  \\


\subsection{Associating ML models with experimental studies}
A long-standing goal of experimental materials characterization of organic optoelectronic materials is a map of how electronic properties are distributed in space as a result of different instantaneous molecular configurations. Scanning probe\cite{ODea-MRS:2012, Groves-ACR:2010, Eisfeld-2018} and super-resolution optical measurements\cite{Penwell-NatMater:2017} can provide readouts of electronic properties on length scales below the diffraction limit. Simultaneously, single-particle measurements\cite{Barbara-ACR:2005, Thiessen-PNAS:2013} can provide a bottom-up understanding of how optoelectronic properties evolve from molecular precursors. For conductive polymers like PEDOT:PSS, single-particle measurements have been particularly difficult to employ due to the lack of emission in these materials because of rapid quenching. Simultaneously, single-particle measurements have tremendous spectroscopic utility due to reduced inhomogeneous broadening. Use of high quality-factor optical microresonators as the readout for ultrasensitive photothermal spectroscopy\cite{Heylman-NatPhoton:2016} has allowed the first single-particle optical measurements to be performed on PEDOT:PSS,\cite{Randall:NL-2018} even down to a single or a small number of polymer strands. This study provided an experimental bound for the line broadening used in the above simulations. More recently, optical microresonator spectroscopy has been used to show how annealing processing act on single PEDOT:PSS polymer particles.\cite{Rea-InReview} A means of directly connecting spectral measurements on single PEDOT:PSS polymer strands and particles to electronic couplings would significantly amplify the information content of these experiments.\\
 

\section{Conclusion}
    Our findings suggest that ML models can identify physical correlations between the measurable electronic absorption spectra and the strength of intermolecular electronic couplings, which in turn determine the charge transport. While the presented ML models provide coupling strength estimates with limited accuracy, relative estimates with respect to reference coupling strengths show promising error rates. Using the trained Bayesian CNN model to classify given electronic absorption spectra above or below an \textit{a priori} selected reference coupling strength displays an error rate of only \unit[12.6]{\%}, and as low as \unit[10]{\%} at high coupling regime. With such a promising error rate, we suggest to use the trained models as classifiers to evaluate performance of fabrications procedure and post-processing steps. Further investigations towards the construction of reliable and transferable ML models, notably the usage of ensemble methods such as adaboost,\cite{Freund:1997,Freund:1996} or mixture density networks,\cite{Bishop:1994,Richter:2003} might allow for more detailed insights into the relation between couplings and electronic absorption spectra. Another important venue for improvement of our approach is the incorporation of features, such as structural information, which would introduce the notion of similarity between complexes.\\
    
	We believe that the combination of the developed approach with spectroscopy techniques and its integration with the self-driving laboratories\cite{SciRobot18:Roch,Tabor:2018,Hase-2019,MI-report,Dimitrov-ACS:2019} has the potential to enhance characterization and accelerate optimization of organic materials. As experimental approaches for providing optical readouts improve in sensitivity,  spatial resolution, and access to different spectral features, growth in theoretical treatments will allow one to draw deeper connections between these measurements and the underlying molecular structure. We also envision the use spectroscopic methods to measure spectra of nanoaggregates with high-finesse toroidal optical cavities.

\begin{acknowledgement}

The computations were done on the Arran cluster supported by the Health Sciences Platform (HSP) at Tianjin University, P.R. China, on the Odyssey cluster supported by the FAS Division of Science, Research Computing Group at Harvard University, USA, and on the computational resource ForHLR II funded by the Ministry of Science, Research and the Arts Baden-W{\"u}rttemberg and DFG (``Deutsche Forschungsgemeinschaft''). L. M. R. and A. A.-G. acknowledge Natural Resources Canada (EIP2-MAT-001) for their financial support. F. H. acknowledges support from the Herchel Smith Graduate Fellowship and the Jacques-Emile Dubois Student Dissertation Fellowship. P. F. has received funding from the European Unions Horizon 2020 research and innovation program under the Marie Skodowska-Curie grant agreement MolDesign No 795206. L. M. R., S. K. S., F. H., P. F., and A. A.-G. thank Dr. Anders Fr{\o}seth for generous support. R. H. G. acknowledges the National Science Foundation (NSF: DMR-1610345). S. L. acknowledges the Real Colegio Complutense in Harvard for a Research Grant, and to the Spanish Ministerio de Ciencia e Innovaci{\'o}n for a Fellowship through the Salvador de Madariaga Program.

\end{acknowledgement}

\begin{suppinfo}
        
The following files are available free of charge.
\begin{itemize}
  \item Supporting\textunderscore Information.pdf: Supporting information to the main manuscript
\end{itemize}        
        
\end{suppinfo}

\newpage
\setcounter{section}{0}

\begin{center}
\Large{\textbf{SUPPORTING INFORMATION}}\\
\Large{\textbf{From absorption spectra to charge transfer in PEDOT nanoaggregates with machine learning}}
\end{center}

\section{Assessment of the basis set}
\label{sec:basis_set}
\begin{table}[htb!]
\centering
\begin{tabular}{l l | c c | c}
\hline\hline
\multicolumn{2}{l}{Pair \#1} & $\epsilon_\mathrm{HOMO}$ & $\epsilon_\mathrm{HOMO-1}$ & $\Delta \epsilon$\\
\hline
\multirow{4}{*}{$\left| \mathrm{mono}_A\right>_{+2}$} & Def2-SVP & -9.134 & -10.018 & 885 \\
& Def2-SVPD & -9.254 & -10.141 & 887 \\
& Def2-SV(P) & -9.144 & -10.030 & 885 \\
& Def2-TZVP & -9.164 & -10.051 & 887 \\
\hline
\multirow{4}{*}{$\left| \mathrm{mono}_A\right>_{+3}$} & Def2-SVP & -11.850 & -12.420 & 570 \\
& Def2-SVPD & -11.934 & -12.505 & 571 \\
& Def2-SV(P) & -11.863 & -12.434 & 571 \\
& Def2-TZVP & -11.842 & -12.411 & 569 \\
\hline
\multirow{4}{*}{$\left| \mathrm{mono}_B\right>_{+2}$} & Def2-SVP & -9.139 & -10.067 & 928 \\
& Def2-SVPD & -9.263 & -10.189 & 926 \\
& Def2-SV(P) & -9.149 & -10.077 & 929 \\
& Def2-TZVP & -9.171 & -10.096 & 925 \\
\hline
\multirow{4}{*}{$\left| \mathrm{mono}_B\right>_{+3}$} & Def2-SVP & -11.932 & -12.518 & 586 \\
& Def2-SVPD & -12.017 & -12.605 & 588 \\
& Def2-SV(P) & -11.944 & -12.531 & 586 \\
& Def2-TZVP & -11.921 & -12.510 & 589 \\
\hline
\multirow{4}{*}{$\left| \mathrm{dimer}\right>_{+5}$} & Def2-SVP & -11.721 & -12.594 & 873 \\
& Def2-SVPD & -11.860 & -12.733 & 873 \\
& Def2-SV(P) & -12.319 & -12.809 & 489 \\
& Def2-TZVP & -12.358 & -12.739 & 381 \\
\hline\hline
\end{tabular}
\caption{Energies of the HOMO and HOMO-1 molecular orbitals of pair \#1, in [eV], at the CAM-B3LYP level of theory. $\Delta \epsilon$ is the energy difference between the HOMO and the HOMO-1, in [meV].}
\label{table:pair1_BSet}
\end{table}

\begin{table}[htb!]
\centering
\begin{tabular}{l l | c c | c}
\hline\hline
\multicolumn{2}{l}{Pair \#2} & $\epsilon_\mathrm{HOMO}$ & $\epsilon_\mathrm{HOMO-1}$ & $\Delta \epsilon$\\
\hline
\multirow{4}{*}{$\left| \mathrm{mono}_A\right>_{+2}$} & Def2-SVP & -9.126 & -10.047 & 921 \\
& Def2-SVPD & -9.246 & -10.166 & 921 \\
& Def2-SV(P) & -9.137 & -10.058 & 922 \\
& Def2-TZVP & -9.156 & -10.074 & 919 \\
\hline
\multirow{4}{*}{$\left| \mathrm{mono}_A\right>_{+3}$} & Def2-SVP & -12.031 & -12.551 & 520 \\
& Def2-SVPD & -12.119 & -12.645 & 525 \\
& Def2-SV(P) & -12.044 & -12.565 & 521 \\
& Def2-TZVP & -11.653 & -12.445 & 792 \\
\hline
\multirow{4}{*}{$\left| \mathrm{mono}_B\right>_{+2}$} & Def2-SVP & -9.190 & -10.206 & 1017 \\
& Def2-SVPD & -9.301 & -10.321 & 1020 \\
& Def2-SV(P) & -9.200 & -10.218 & 1017 \\
& Def2-TZVP & -9.208 & -10.227 & 1018 \\
\hline
\multirow{4}{*}{$\left| \mathrm{mono}_B\right>_{+3}$} & Def2-SVP & -12.031 & -12.668 & 637 \\
& Def2-SVPD & -12.114 & -12.749 & 635 \\
& Def2-SV(P) & -12.044 & -12.682 & 638 \\
& Def2-TZVP & -12.021 & -12.654 & 633 \\
\hline
\multirow{4}{*}{$\left| \mathrm{dimer}\right>_{+5}$} & Def2-SVP & -11.783 & -12.744 & 961 \\
& Def2-SVPD & -- & -- & -- \\
& Def2-SV(P) & -11.793 & -12.755 & 962 \\
& Def2-TZVP & -11.842 & -12.789 & 947 \\
\hline\hline
\end{tabular}
\caption{Energies of the HOMO and HOMO-1 molecular orbitals of pair \#2, in [eV], at the CAM-B3LYP level of theory. $\Delta \epsilon$ is the energy difference between the HOMO and the HOMO-1, in [meV].}
\label{table:pair2_BSet}
\end{table}

\begin{table}[htb!]
\centering
\begin{tabular}{l | c c}
\hline\hline
J & Pair \#1 & Pair \#2 \\ 
\hline
Def2-SVP & 870 & 959 \\
Def2-SVPD & 870 & -- \\
Def2-SV(P) & 483 & 960 \\
Def2-TZVP & 374 & 893 \\
\hline\hline
\end{tabular}
\caption{J in [meV].}
\label{table:J_BSet}
\end{table}

    \section{Molecular dynamics}
	\label{sec:si_md}
	To ensure equilibration of the systems, the following approach has been adopted. First, random starting configurations have been generated with the packmol program, and each one has been allowed to relax with an energy minimization followed by a molecular dynamics simulation of \unit[1]{ns} within the NVT ensemble at \unit[300]{K}. After that, a NPT simulation of \unit[2]{ns} at \unit[300]{K} and \unit[1]{atm} to relax the dimensions of the box. Then, the temperature of the system has been subjected to a heating up to \unit[1000]{K} in a \unit[1]{ns} run, a NVT run at \unit[1000]{K} for another \unit[1]{ns}, and a cooling down to \unit[300]{K} for \unit[10]{ns}. After that, the production run has been carried out in the NPT ensemble at \unit[300]{K} and \unit[1]{atm} for \unit[1]{ns}. To more thoroughly explore the configurational space, a total of ten different starting configurations generated by Packmol have been simulated through this procedure.

	\section{Hyperparameter optimization of correlation models}
	\label{sec:si_hyperparameter_optimization}
    
        Well performing model architectures were determined from a random grid search of selected hyperparameter for each model. Tab.~\ref{tab:hyperparam_ranges} summarizes the hyperparameters which were varied for each model and their respective ranges. Multi-layer perceptrons (applied to the full spectra and PCA contracted spectra) were set up with three hidden layers but varying number of neurons per layer. Activations for all model architectures were chosen to be either a leaky version of the ReLU function ($\alpha = 0.2$) or the softplus function.
        
        \begin{align*}
            \text{leaky\_ReLU}(x) &= \begin{cases} x, & \text{if } x > 0 \\ \alpha x, & \text{otherwise} \end{cases}  \\
            \text{softplus}(x)    &= \log(1 + \exp(x))
        \end{align*}
        
        Tab.~\ref{tab:hyperparam_ranges} also reports the set of hyperparameters for which the trained models achieved the highest prediction accuracies as determined from a 10-fold cross-validation protocol. Best performing sets of hyperparameters were determined from a random grid search with a total of 512 different models constructed for each model class. Training of individual models was aborted based on an early stopping criterion. 
        
        \begin{table}[!ht]
            \centering
            \begin{tabular}{p{6cm} | p{3cm} | p{2.5cm}p{1.5cm}p{1.5cm}}
                    Model    & Hyperparameter & optimal & lower bound & upper bound \\
                \hline
                    \breg    & Regularization & $10^{-3.875}$ & $10^{-4}$ & $10^1$    \\
                             & Learning rate  & $10^{-4.75}$  & $10^{-5}$ & $10^{-1}$ \\
                \hline
                    \convnet & Regularization    & $10^0$        & $10^{-4}$ & $10^1$    \\
                             & Learning rate     & $10^{-1.875}$ & $10^{-5}$ & $10^{-1}$ \\
                             & Filters           & $24$          & & \\
                             & Filter sizes      & $20, 50$      & & \\
                             & Filter activation & softplus      & \multicolumn{2}{c}{softplus or leaky ReLU} \\
                             & Neurons / layer   & $2$           & & \\
                             & Dense activation  & leaky ReLU    & \multicolumn{2}{c}{softplus or leaky ReLU} \\
                \hline 
                    \mlpf    & Regularization    & $10^{-2}$     & $10^{-4}$ & $10^1$    \\
                             & Learning rate     & $10^{-3.5}$   & $10^{-5}$ & $10^{-1}$ \\
                             & Neurons / layer   & $91$          & $1$       & $100$     \\
                             & Activations       & softplus      & \multicolumn{2}{c}{softplus or leaky ReLU} \\
                \hline 
                    \mlpp    & Regularization    & $10^{-3.5}$   & $10^{-4}$ & $10^1$    \\
                             & Learning rate     & $10^{-1.5}$   & $10^{-5}$ & $10^{-1}$ \\
                             & Neurons / layer   & $41$          & $1$       & $100$     \\
                             & Activations       & leaky ReLU    & \multicolumn{2}{c}{softplus or leaky ReLU} \\
            \end{tabular}
            \caption{Hyperparameter ranges and sets of hyperparameters for which trained models achieved the highest prediction accuracies. Best performing sets of hyperparameters were determined from a random grid search. }
            \label{tab:hyperparam_ranges}
        \end{table}

    \section{Sampling efficiency}
    \label{sec:si_sampling_efficiency}
    
\begin{figure}[!ht]
            \centering
            \includegraphics[width = 1.0\textwidth]{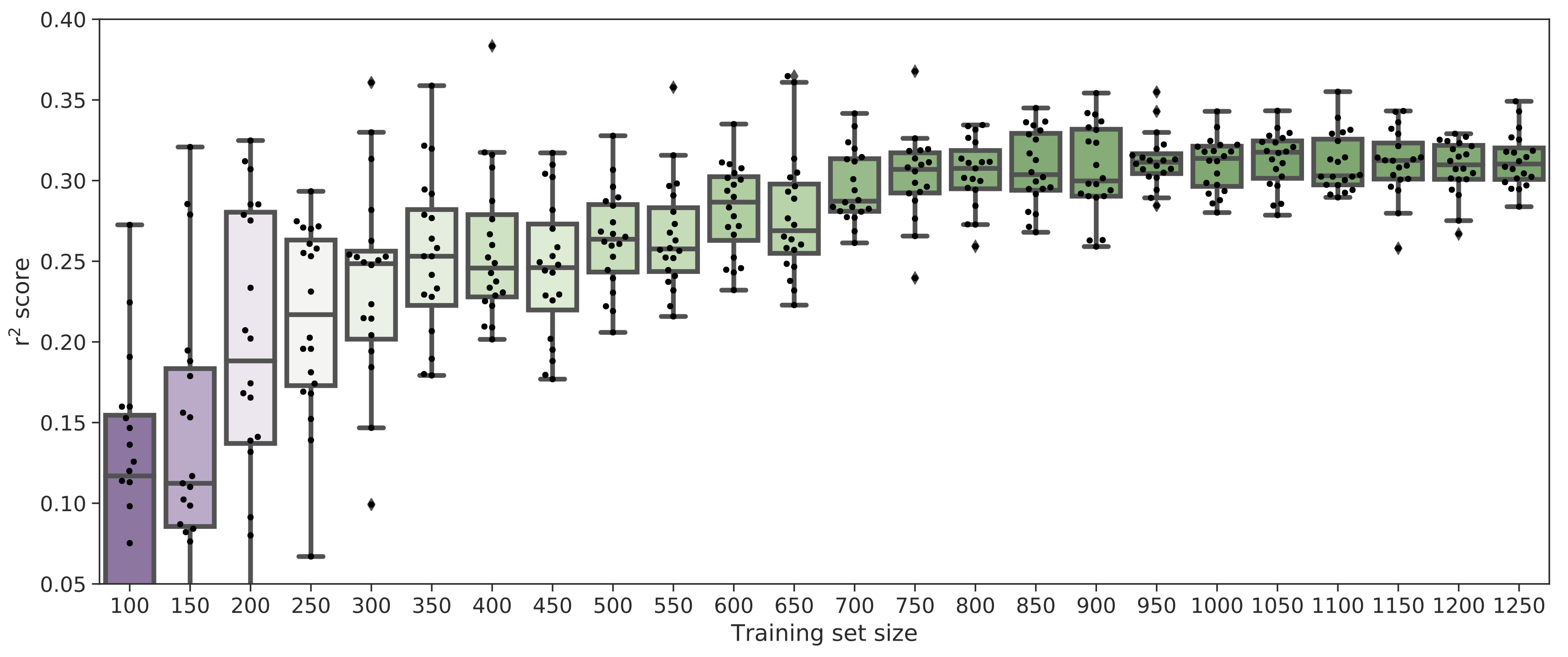}
            \caption{Sampling efficiency for Bayesian convolutional neural networks trained to predict tight-binding coupling strengths from Vis/NIR spectra. Colors indicate poor predictions (purple) and accurate predictions (green). }
            \label{fig:sampling_efficiency}
        \end{figure}

        The predictive power of a regression model depends on the how representative the training set is for the underlying (unknown) physical correlations. Typically, larger training sets with more examples of absorption spectra and associated couplings yield higher prediction accuracies. To asses how many examples are needed for sufficient prediction accuracies we train the introduced regression models on training sets of different sizes. More specifically, we employ the \convnet model (see main text for details) with the determined best performing set of hyperparameters. The training set of 1,250 examples is reduced by randomly selected examples. Following this procedure, we generate a total of $20$ training subsets of different sizes. \convnet models are then fully trained on their respective training set and their predictive powers are assessed via test set predictions. Note, that the test set is identical for all predictions. Results are reported in Fig.~\ref{fig:sampling_efficiency}
    
        We observe a steady increase in the average prediction accuracy of the \convnet model when increasing the size of the training set from $100$ examples up to about $850$ examples. From there on, no significant change in the average prediction accuracy can be observed with a further increase of the training set size. 
        
    \section{Prediction accuracies for coupling strengths computed with the Kohn-Sham orbital formalism}
    \label{sec:si_cross_predictions}
    
        ML models can, at best, identify statistical correlations between the features and the targets to which they are exposed during the training process. As such, the positive correlation between predicted and computed coupling strengths obtained from the tight-binding formalism does not necessarily indicate that the ML models identified relevant physical correlations, and might just be due to statistical correlations based on the impreciseness of the methods used to compute Vis/NIR spectra and/or coupling strengths. \\
        
        We suggest to test if the ML models in the main text only identified statistical correlations (and not the relevant physical correlations) by constructing additional models trained to predict coupling strengths computed with the Kohn-Sham orbital formalism. Similar to the ML models reported in the main text, we run a full hyperparameter optimization for the Bayesian CNN models to predict Kohn-Sham orbital formalism coupling strengths. Test set predictions of the best performing model as identified from the hyperparameter search are reported in Fig.~\ref{fig:sampling_efficiency}. We observe, that the trained Bayesian CNN is capable of identifying correlations between the Vis/NIR spectra and the coupling strengths even for this case where coupling strengths are computed based on the Kohn-Sham orbital formalism. Moreover, we observe similar prediction accuracies despite significantly different coupling strength distributions (see section \ref{sec:si_coupling_strength_distributions}).
        
        \begin{figure}[!ht]
            \centering
            \includegraphics[width = 0.5\textwidth]{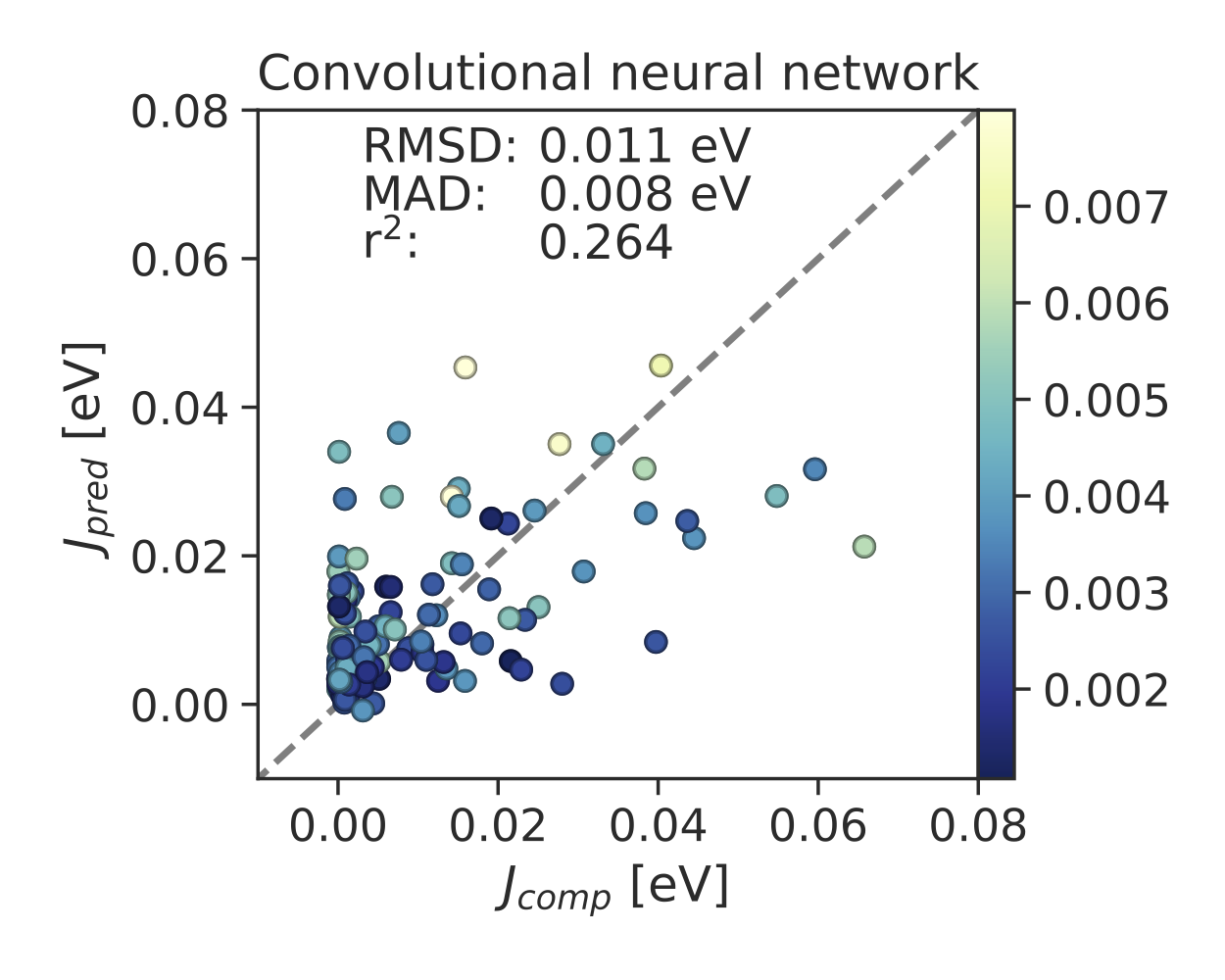}
            \caption{Predictions for coupling strengths computed based on the Kohn-Sham orbital formalism. Coupling strengths are predicted from a Bayesian convolutional neural network for which hyperparameters have been optimized in a random grid search. }
            \label{fig:kohn_sham_predictions}
        \end{figure}
        
        If the predictive power of the Bayesian CNN reported in the main text relied on spurious statistical correlations between the Vis/NIR spectra and the tight-binding coupling strengths, the model presented in this section no longer has the opportunity to exploit these spurious correlations due to the change in formalism for computing coupling strengths. However, we still observe relatively high prediction accuracies for Kohn-Sham orbital couplings despite the change in the range of the couplings and the change in their distribution. We conclude that the predictive power of the Bayesian CNN model presented in Fig.~\ref{fig:kohn_sham_predictions} must either be due to the model identifying the relevant physical correlations, or other statistical correlations arising from the Kohn-Sham orbital couplings. \\
        
        To rule out the possibility that the Bayesian CNN presented in Fig.~\ref{fig:kohn_sham_predictions} identified other spurious statistical correlations yielding similar prediction accuracies we propose the construction of a hybrid dataset. This hybrid dataset is constructed by randomly choosing half of the coupling strengths computed with the tight-binding formalism, and the other half with the Kohn-Sham orbital formalism. Note, that the differences in the ranges of the coupling strengths are accounted for by standardizing coupling strengths based on the characteristics of each individual dataset following
        
        \begin{align}
            j = \frac{J - \langle J \rangle_\text{train}}{\sqrt{ \langle \left( J - \langle J \rangle_\text{train} \right)^2 \rangle_\text{train} }},
        \end{align}
        
        where $\langle \cdot \rangle_\text{train}$ denotes the average over the training set.
        
        \begin{figure}[!ht]
            \centering
            \includegraphics[width = 0.5\textwidth]{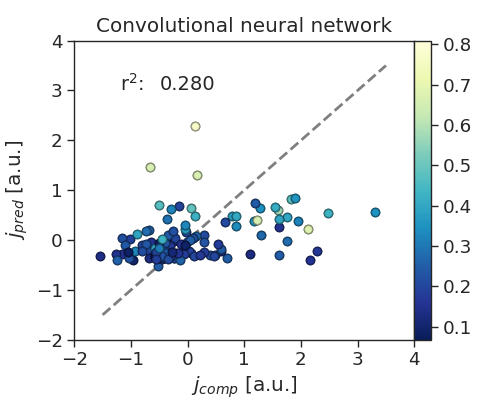}
            \caption{Coupling strength predictions obtained from a model trained on a hybrid dataset assembled from both formalisms (equal contributions). Note, that coupling strengths have been standardized before training the model, and the model predicted coupling strengths in standardized units (shown in the plot). }
            \label{fig:kohn_sham_predictions}
        \end{figure}
        
        We find that the proposed Bayesian CNN architecture yields prediction accuracies comparable to prior experiments with coupling strengths computed from one or the other formalism. We interpret this observation as an indicator showing that the model does not solely rely on spurious statistical correlations caused by inaccuracies of the physical models used to compute the coupling strengths.

    \section{Distribution of electronic coupling}
    \label{sec:si_coupling_strength_distributions}
        \begin{figure}[htb!]
         \centering
         \includegraphics[width=0.7\linewidth]{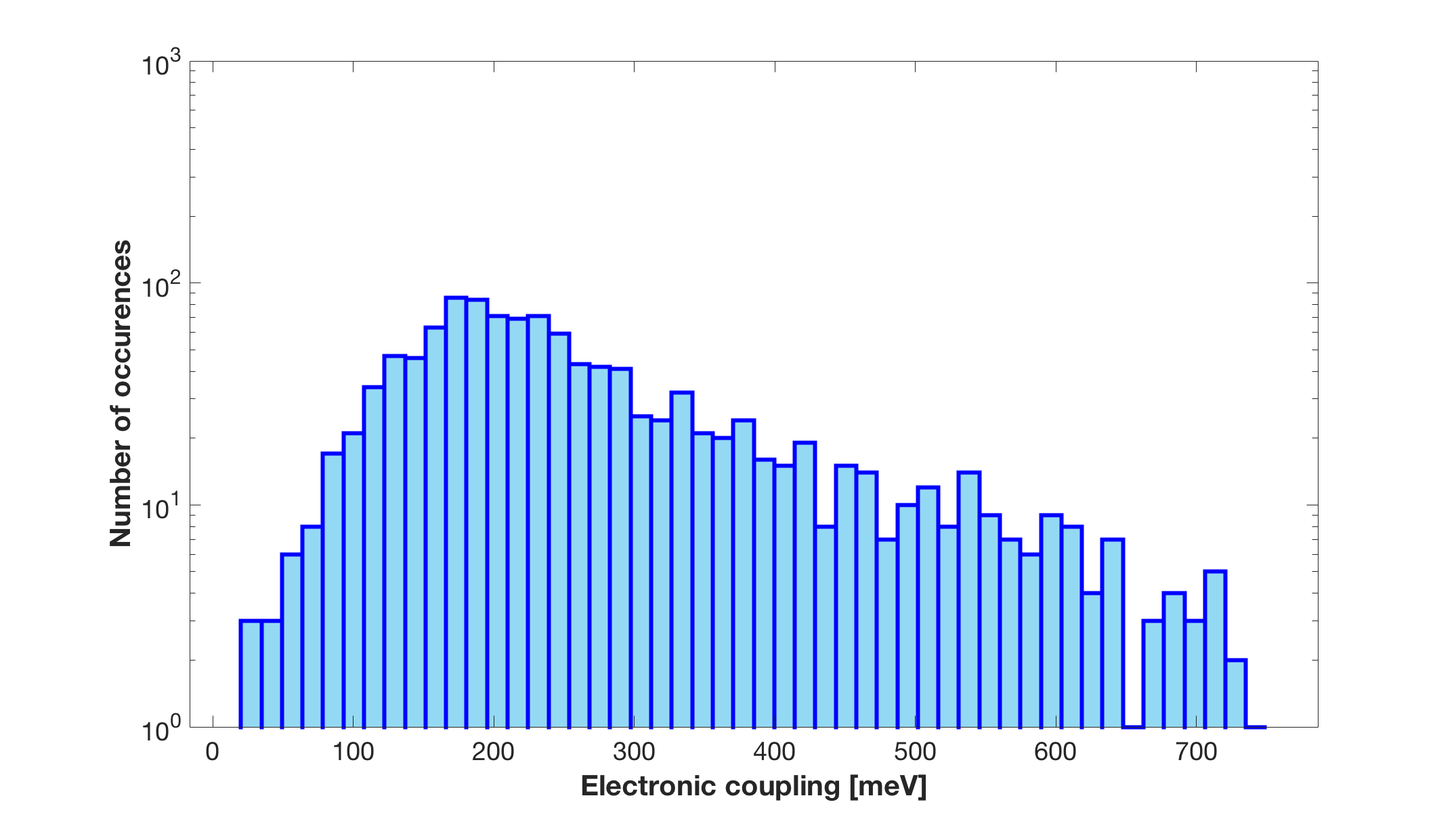}
         \includegraphics[width=0.7\linewidth]{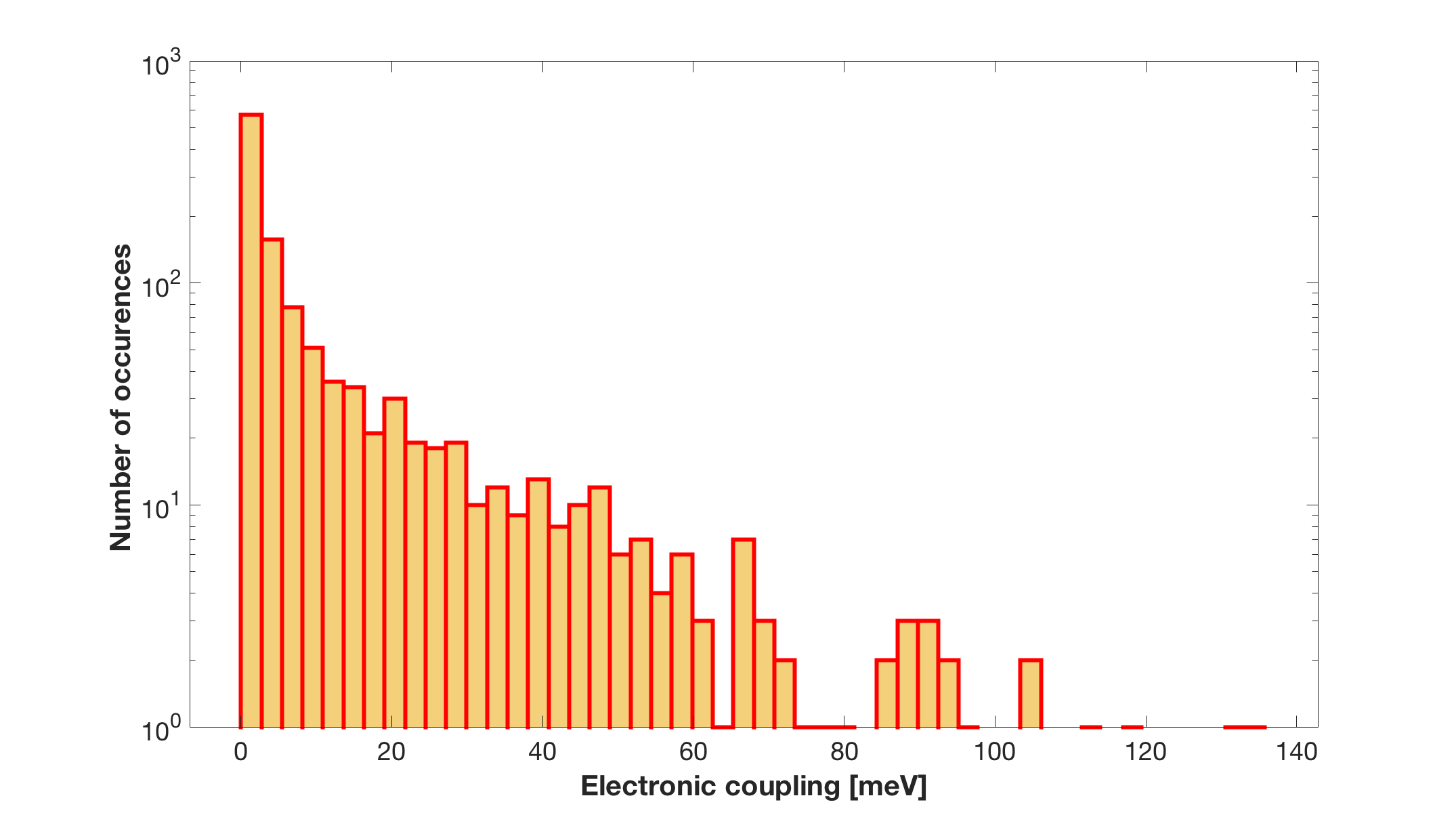}
        \caption{Distribution of electronic couplings in PEDOT:PEDOT complexes. The total number of complexes is 1,420. (A) is computed with the tight binding formalism model, and (B) with the orbital overlap method.}
        \label{fig:couplings}
        \end{figure}


\bibliography{main.bib}

\end{document}